\documentclass[letterpaper,twocolumn,10pt]{article}
\usepackage{usenix2019_v3}

\usepackage{tikz}
\usepackage{amsmath,amssymb,amsfonts}
\usepackage{tabu}
\usepackage{subfig}   
\usepackage{algorithm}
\usepackage{algorithmic}
\usepackage{filecontents}
\usepackage{titling}
\usepackage{comment}
\usepackage{authblk}
\usepackage[bottom]{footmisc}

\begin{document}

\date{}

\setlength{\droptitle}{-3em}
\title{\Large \bf Metaflow: A DAG-Based Network Abstraction \\ for Distributed Applications}

\author[1]{Jiawei Fei \thanks{Jiawei Fei and Yang Shi are students.}}
\author[1]{Yang Shi}
\author[2]{Qun Huang}
\author[1]{Mei Wen}
\affil[1]{College of Computer Science and Technology, National University of Defense Technology \authorcr Email: \{feijiawei11, shiyang14, wenmei\}@nudt.edu.cn}
\affil[2]{Institute of Computing Technology, Chinese Academy of Sciences, \authorcr Email: huangqun@ict.ac.cn }

\renewcommand\Authands{ and }

\maketitle


\section{Introduction}
The purpose of network optimization is to boost the distributed application performance. 
In the past decade, increasingly network scheduling techniques have been proposed to achieve this goal which focus on network metrics in different levels. 
Flow-level metrics, such as flow completion time (FCT) and per-flow fairness, are based on the abstraction of flows yet they cannot capture the semantics of communication in a cluster application. 
Therefore, flow-level metrics can be at odds with application-level goals.
Being aware of this problem, coflow\cite{chowdhury2012coflow} is proposed as a new network abstraction.
It is a major leap forward of application-aware network scheduling.
Minimizing the average coflow completion time (CCT) usually aligns application-level performance, thereby actually decreasing job completion time (JCT).
However, the coflow abstraction is insufficient to reveal the dependencies between computation and communication.
As a result, the real application performance can be hurt, especially in the absence of hard barriers. 

\begin{figure}[!t]
	\centering
	\subfloat[Scheduling problem over a 3$ \times $3 DC fabric with three ingress/egress ports.]{
		\label{Fig:motivation-a}
		\begin{minipage}[t]{0.5\linewidth}
			\centering
			\includegraphics[width=1.7in]{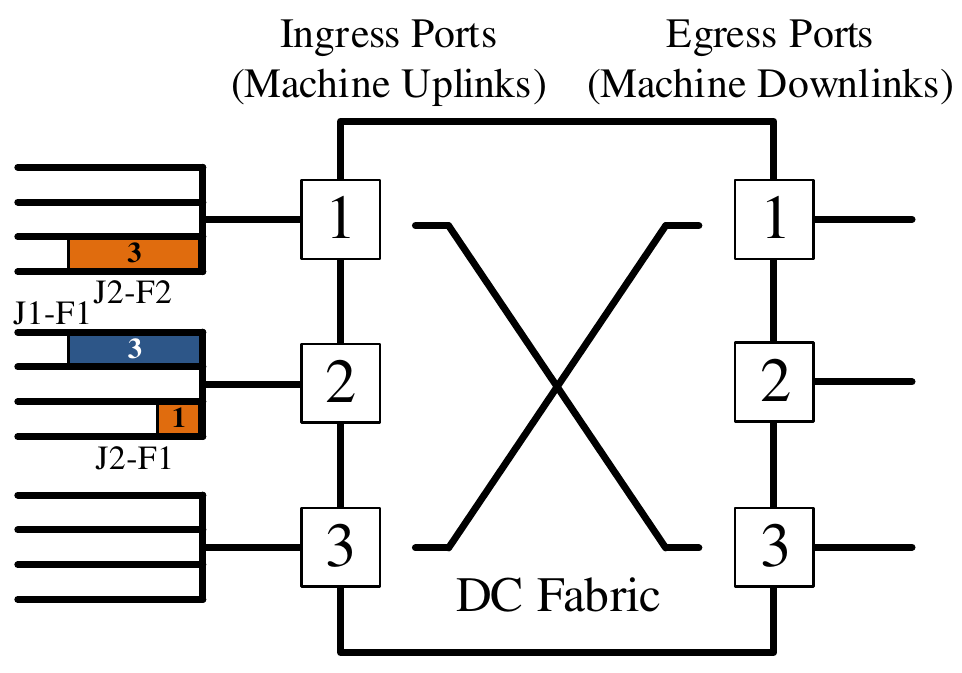}
		\end{minipage}
	}
	\subfloat[Computation DAG of two jobs.]{
		\label{Fig:motivation-b}
		\begin{minipage}[t]{0.5\linewidth}
			\centering
			\includegraphics[width=0.85in]{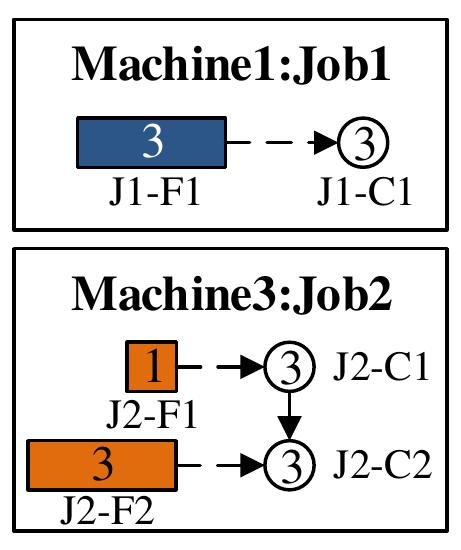}
		\end{minipage}
	}\\
	\subfloat[CCT-Optimized]{
		\label{Fig:motivation-c}
		\begin{minipage}[t]{0.5\linewidth}
			\centering
			\includegraphics[width=1.8in]{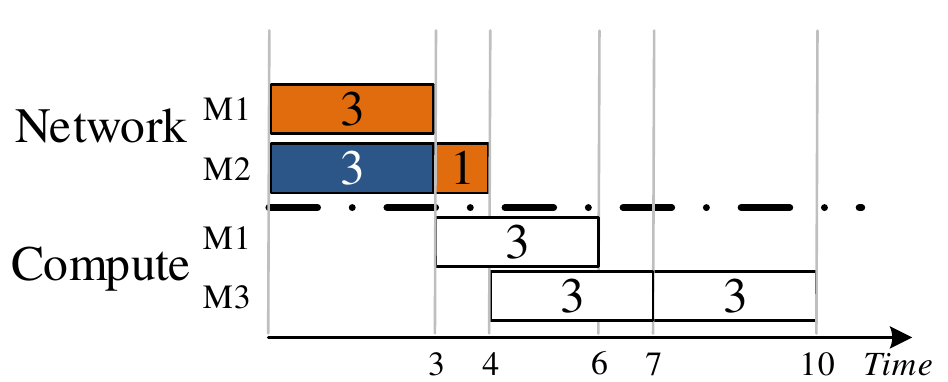}
		\end{minipage}
	}
	\subfloat[JCT-Optimized.]{
		\label{Fig:motivation-d}
		\begin{minipage}[t]{0.5\linewidth}
			\centering
			\includegraphics[width=1.1in]{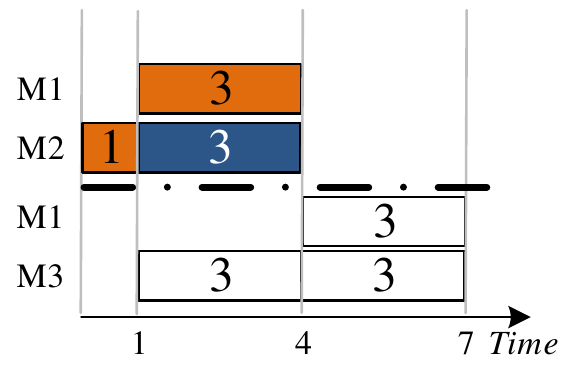}
		\end{minipage}
	}
	\caption{A motivation example}
	\vspace{-0.7cm}
	\label{Fig:motivation}
\end{figure}

For example, a scheduling problem is shown in Figure~\ref{Fig:motivation}. 
In this case, there are two jobs J1 and J2 and every job has a coflow (C1 and C2). 
C1 has one flow transferring 3 units of data from machine 2 to machine 1, while C2 has two flows transferring 3 and 1 units of data from machine 1 and machin 2 to machine 3. 
Additionally, the subsequent computation of J1 and J2 is performed in machine 1 and machine 3 respectively. 
The dependencies between flows and computation are discribed with DAGs shown in Figure~\ref{Fig:motivation-b}. 
To minimize average CCT, the optimal schedule is shown in Figure~\ref{Fig:motivation-c} and the average CCT is just $\frac{3+4}{2}=3.5$ units of time. 
Besides, based on DAGs of jobs, the average JCT can be calculated which is $\frac{6+10}{2}=8$ units of time.

As a comparison, Figure~\ref{Fig:motivation-d} shows a different scheduling.
Obviously, it introduces more overlaps between communication and computation.
Thus, the average JCT is reduced to $\frac{7+7}{2}=7$ units while the CCT ($\frac{4+4}{2}=4$) is higher. 

In this paper, we propose a new abstraction namely \textit{metaflow} that resides in the middle of the two extreme points of flows and coflows.
Each metaflow is a collection of flows consumed by the same computation task in the DAG of one job.
The job is assumed to be executed in a distributed environment in data level parallelism manner.

\begin{figure}[]
	\centering
	\subfloat[A job.]{
		\label{Fig:def1}
		\begin{minipage}[t]{0.5\linewidth}
			\centering
			\includegraphics[width=1.7in]{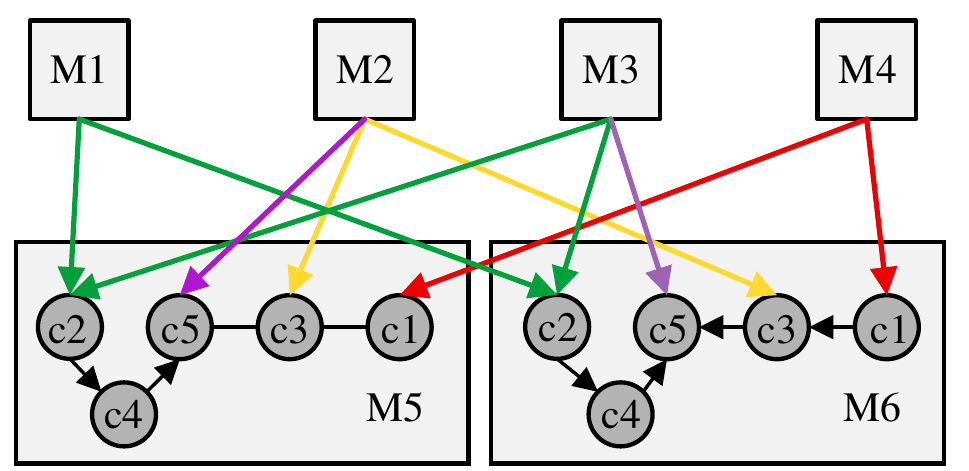}
		\end{minipage}
	}
	\subfloat[Metaflows.]{
		\label{Fig:def2}
		\begin{minipage}[t]{0.5\linewidth}
			\centering
			\includegraphics[width=1.0in]{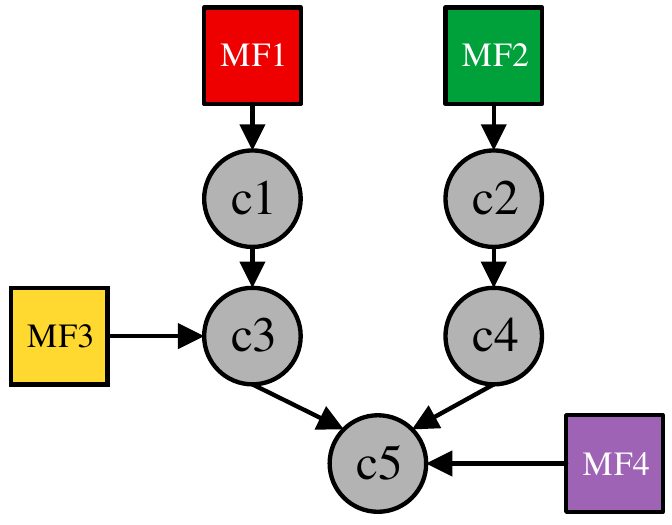}
		\end{minipage}
	}
	\caption{A metaflow example}
	\vspace{-0.7cm}
	\label{Fig:definition}
\end{figure}

Considering an example job in Figure~\ref{Fig:def1} with 4 sending and 2 receiving machines, all flows are classified into one coflow.
Meanwhile, following the definition of metaflow, these flows are divided into 4 metaflows, denoted with different colors.
Each metaflow is connected to an exclusive computing task, as shown in Figure~\ref{Fig:def2}.
Regarding the DAG, a metaflow is the smallest unit to forward the computation progress.

\section{Scheduling Algorithms}
We propose a \underline{\textbf{m}etaflow-based} \textbf{s}cheduling \textbf{a}lgorithm (MSA) to schedule the metaflows, as detailed in Algorithm~\ref{alg:alg1}.
It has three main steps:
firstly, we calculate the performance gain brought by each metaflow;
then, all active metaflows are sorted according to their gains in decreasing order;
at last, the available bandwidth is assigned to each metaflow in order.

\textbf{Performance gain estimation.}
Metaflow's gain can be classified into two categories, \textit{direct} and \textit{indirect} gain, according to whether this metaflow can invoke the computation independently.
If it could, the gain of this metaflow is calculated in two steps: first the computation load of these tasks are summed up, and then divided by the remained size of the metaflow.
For example, the metaflow 1 and 2 in Figure~\ref{Fig:def2} both can result subsequent computing tasks dependently.
The profits of them are $load_{c1}/reSize_{MF1}$ and $(load_{c2}+load_{c4})/reSize_{MF2}$ respectively.

In the other case, a metaflow can not start a computing tasks dependently, that is to say, it has to wait other unfinished metaflows.
For this kind of metaflows, the attributes are set to the total metaflow sizes needed to start the computing task.
In Figure~\ref{Fig:def2}, metaflow 3 and 4 belong to this kind, and the attributes of them are $(reSize_{MF1}+reSize_{MF3})$ and $(reSize_{MF1}+reSize_{MF2}+reSize_{MF3}+reSize_{MF4})$.

\textbf{Sorting.}
Metaflows which have direct profits are superior to that which have indirect profits.
In their individual group, metaflows with larger direct profit and smaller indirect profit are given higher priorities.

\textbf{Bandwidth assignment.}
For a selected metaflow, it contains flows go to all reducers of this job.
Since the JCT is the maximum finish time of these reducers, all reducers are expected to finish simultaneously.
Some existing algorithms have been proposed to this intent, say MADD~\cite{chowdhury2014efficient}, we can adopt these algorithms directly.
After assign bandwidth to a metaflow, the available bandwidth in this time slot is updated and the assignment of next metaflow begins.
This assign process ends until all metaflows are considered or no bandwidth resource is left.

\begin{algorithm}[]
	\caption{Metaflow Scheduling Algorithm (MSA)}
	\label{alg:alg1}
	\begin{algorithmic}[1]
		\IF{(metaflow arrives \textbf{or} finishes)}
		\IF{no unfinished metaflows}
		\STATE {\textbf{break}}
		\ENDIF 
		\STATE {extract all unfinished metaflows}
		\FOR {unfinished $metaflow_i$}
		\STATE {calculate the gain of $metaflow_i$}
		\ENDFOR
		\STATE {sort metaflows based on profit}
		\FOR {$metaflow_i$ in the sorted list}
		\STATE {assign bandwidth to $metaflow_i$ (e.g. MADD)}
		\ENDFOR
		\ENDIF
	\end{algorithmic}
\end{algorithm}

\section{Preliminary Results}
\begin{figure}[!t]
	\centering
	\subfloat[DAGs with different topologies]{
		\label{Fig:result1}
		\begin{minipage}[t]{0.5\linewidth}
			\centering
			\includegraphics[width=1.3in]{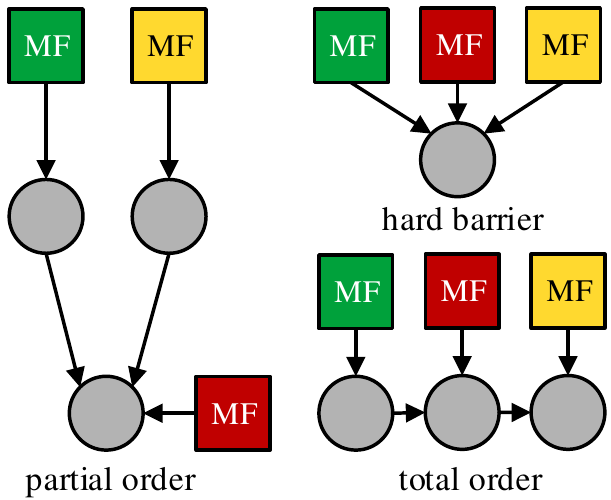}
		\end{minipage}
	}
	\subfloat[JCT and speedup]{
		\label{Fig:result2}
		\begin{minipage}[t]{0.5\linewidth}
			\centering
			\includegraphics[width=1.65in]{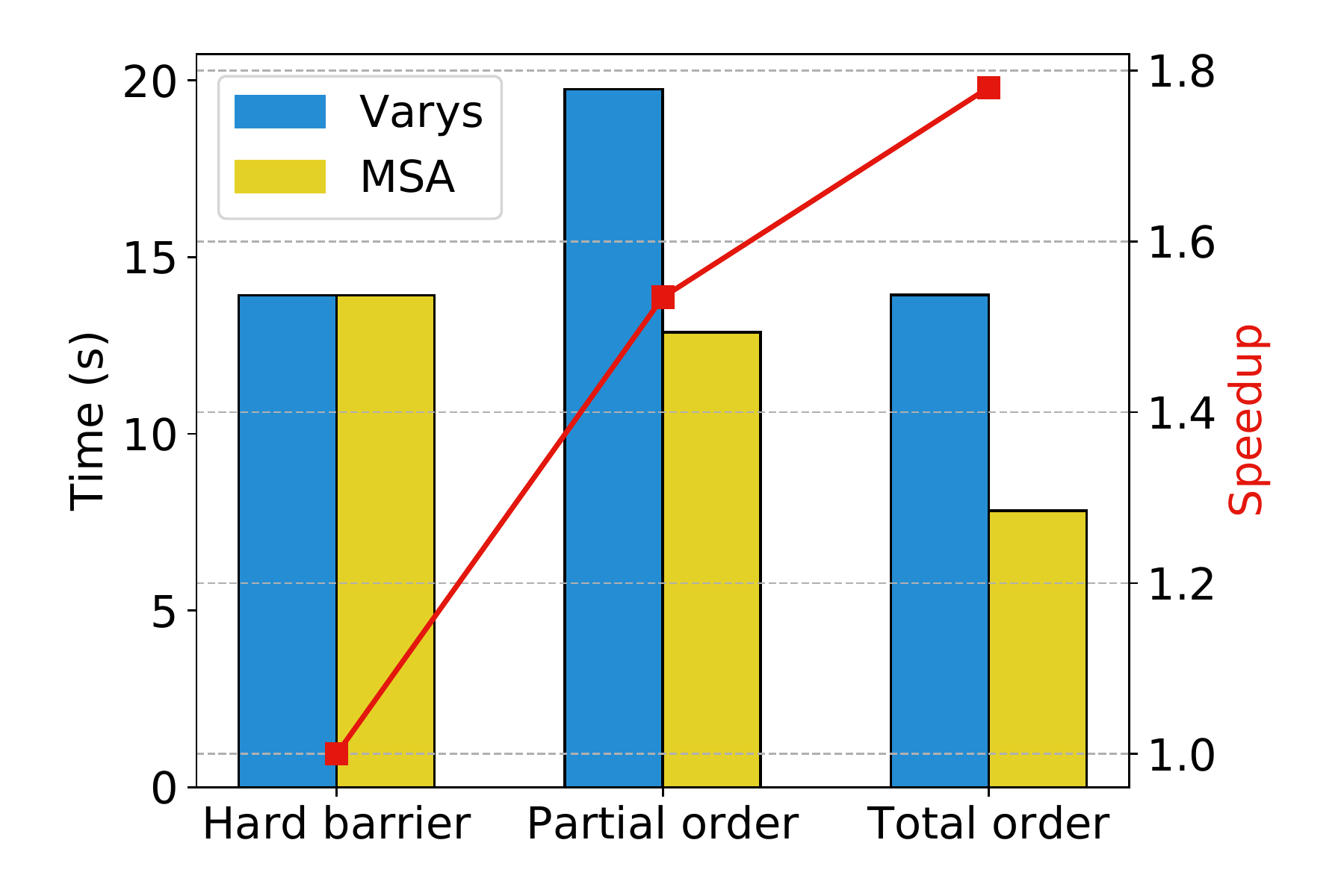}
		\end{minipage}
	}
	\caption{DAG topologies and evaluation results}
	\label{Fig:results}
\end{figure}

In this section, we evaluate the performance of MSA for a single job using a flow-level simulator.
We take the collected Facebook logs~\cite{chowdhury2014efficient} as our workload.
The Facebook logs lack the DAG information of jobs, thus we generate the DAGs for each job.
Figure~\ref{Fig:result1} illustrates three types of topologies we use, which are partial order, disorder and total order respectively.

For each topology, we randomly select 50 jobs from the workload and compute the average JCT as results.
We compare MSA with a coflow-based scheduling algorithm (Varys)~\cite{chowdhury2014efficient}.
As shown in Figure~\ref{Fig:result2}, MSA outperforms Varys by 1.78$\times$ and 1.53$\times$ for DAGs in total order and partial order separately.
Even in presence of the hard barrier, MSA is equivalent to Varys and achieves the same JCT.
In summary, metaflow-based scheduling can effectively improve the distributed application performance.

\section{Ongoing Work}
As part of our current work, we are trying to deploy metaflow into real clusters.
Our end goal is to design a self-driven system which automatically detects and schedules the metaflows to optimize the applications' performance.
As a first step, we plan to devise a learning based mechanism that can distinguish metaflows along with their attributes using collected network statistics.
Further, the routing strategies of metaflows will be investigated, which is an indispensable component in real network.

\bibliographystyle{plain}
\bibliography{\jobname}

\begin{thebibliography}{1}

\bibitem{chowdhury2012coflow}
Mosharaf Chowdhury and Ion Stoica.
\newblock Coflow: A networking abstraction for cluster applications.
\newblock In {\em Proceedings of the 11th ACM Workshop on Hot Topics in
  Networks}, pages 31--36. ACM, 2012.

\bibitem{chowdhury2014efficient}
Mosharaf Chowdhury, Yuan Zhong, and Ion Stoica.
\newblock Efficient coflow scheduling with varys.
\newblock In {\em ACM SIGCOMM Computer Communication Review}, volume~44, pages
  443--454. ACM, 2014.

\end{thebibliography}

\end{document}